\documentclass[]{spie}  

\usepackage[]{graphicx}

\usepackage{color}
\definecolor{blue}{rgb}{0.1,0.1,0.6}
\definecolor{orange}{rgb}{0.74,.35,0.099}
\definecolor{pale}{rgb}{0.90,0.90,0.95}
\definecolor{red}{rgb}{1.0,0.0,0.0}

\newcommand{\refnum}[1]{Ref.~\citenum{#1}}

\title{Gemini Planet Imager Observational Calibrations I: \\Overview of the GPI Data Reduction Pipeline}

\author{Marshall D. Perrin\supit{a}, 
{Jerome Maire}\supit{b},
{Patrick Ingraham}\supit{c,d},
{Dmitry Savransky}\supit{e},\\
{Max Millar-Blanchaer}\supit{f}, 
{Schuyler G. Wolff}\supit{g,a},
{Jean-Baptiste Ruffio}\supit{h},
{Jason J. Wang}\supit{i},\\
{Zachary H. Draper}\supit{j,k},
{Naru Sadakuni}\supit{l},
{Christian Marois}\supit{k,j},
{Abhijith Rajan}\supit{m},
Michael P. Fitzgerald\supit{n}, 
Bruce Macintosh\supit{c,o}, 
James R. Graham\supit{i}, 
{Ren\'e Doyon}\supit{d},
{James E. Larkin}\supit{n},
{Jeffrey K. Chilcote}\supit{n},
{Stephen J. Goodsell}\supit{l},
{David W. Palmer}\supit{o},
{Kathleen Labrie}\supit{p},
{Mathilde Beaulieu}\supit{q},
{Robert J. De Rosa}\supit{m,r},
{Alexandra Z. Greenbaum}\supit{g,a},
{Markus Hartung}\supit{l},
{Pascale Hibon}\supit{l},
{Quinn Konapacky}\supit{b},
{David Lafreniere}\supit{d},
{Jean-Francois Lavigne}\supit{s},
{Franck Marchis}\supit{h},
{Jenny Patience}\supit{m},
{Laurent Pueyo}\supit{a},
{Fredrik T. Rantakyr\"o}\supit{l},
{R\'emi Soummer}\supit{a},
{Anand Sivaramakrishnan}\supit{a},\\
{Sandrine Thomas}\supit{t,u},
{Kimberly Ward-Duong}\supit{m}, and
{Sloane Wiktorowicz}\supit{v}
\skiplinehalf
\supit{a} Space Telescope Science Institute, 3700 San Martin Dr, Baltimore, MD 21218, USA \\
\supit{b} Dunlap Institute for Astrophysics, University of Toronto, 
50 St. George St, Toronto ON M5S 3H4, Canada
\supit{c} Kavli Institute for Particle Astrophysics and Cosmology, Stanford University, Stanford, CA 94305, USA
\supit{d} Department de Physique, Universit\'{e} de Montr{\'e}al, Montr\'eal QC H3C 3J7, Canada
\supit{e} Sibley School of Mechanical and Aerospace Engineering, Cornell University, Ithaca NY 14853, USA
\supit{f} Department of Astronomy \& Astrophysics, University of Toronto, Toronto ON M5S 3H4, Canada
\supit{g} Physics \& Astronomy Department, Johns Hopkins University, Baltimore MD, 21218, USA
\supit{h} SETI Institute, Carl Sagan Center, 189 Bernardo Avenue, Mountain View, CA 94043, USA
\supit{i} Department of Astronomy, UC Berkeley, Berkeley CA, 94720, USA
\supit{j} University of Victoria, 3800 Finnerty Road Victoria BC V8P 5C2, Canada
\supit{k} National Research Council of Canada Herzberg, 5071 West Saanich Road, Victoria, BC V9E 2E7, Canada
\supit{l} Gemini Observatory, Casilla 603, La Serena, Chile
\supit{m} School of Earth and Space Exploration, Arizona State University, 
PO Box 871404, Tempe, AZ 85287, USA 
\supit{n} Department of Physics and Astronomy, UCLA, Los Angeles, CA 90095, USA
\supit{o} Lawrence Livermore National Lab, 7000 East Ave., Livermore, CA 94551, USA
\supit{p} Gemini Observatory, 670 N. A'ohoku Place, Hilo, Hawaii, 96720, USA
\supit{q} Universit\'e de Nice Sophia-Antipolis, Observatoire de la C\^ote d'Azur, CNRS UMR 7293, 06108 Nice Cedex 2, France
\supit{r} School of Physics, College of Engineering, Mathematics and Physical Sciences, University of Exeter, Stocker Road, Exeter, EX4 4QL, UK
\supit{s} ABB, Inc. Quebec, Canada.
\supit{t} NASA Ames Research Center,  Mountain View, CA 94035, USA
\supit{u} UARC, UC Santa Cruz, Santa Cruz CA 95065, USA
\supit{u} UC Santa Cruz, 1156 High Street, Santa Cruz, CA 95064, USA
}

\authorinfo{Please send correspondence to M.D.P. at mperrin@stsci.edu.}

\begin{document} 
  \maketitle 

\begin{abstract}
The Gemini Planet Imager (GPI) has as its science instrument an infrared integral field spectrograph/polarimeter (IFS).  Integral field spectrographs are scientificially powerful but require sophisticated data reduction systems. For GPI to achieve its scientific goals of exoplanet and disk characterization, IFS data must be reconstructed into high quality astrometrically and photometrically accurate datacubes in both spectral and polarization modes, via flexible software that is usable by the broad Gemini community.  The data reduction pipeline developed by the GPI instrument team to meet these needs is now publicly available following GPI's commissioning.  

This paper, the first of a series, provides a broad overview of GPI data reduction, summarizes key steps, and presents the overall software framework and implementation.
Subsequent papers describe in more detail the algorithms necessary for calibrating GPI data.
The GPI data reduction pipeline is open source, available from \texttt{planetimager.org}, and will continue to be enhanced throughout the life of the instrument. It implements an extensive suite of task primitives that can be assembled into reduction recipes to produce calibrated datasets ready for scientific analysis.  Angular, spectral, and polarimetric differential imaging are supported. Graphical tools automate the production and editing of recipes, an integrated calibration database manages reference files, and an interactive data viewer customized for high contrast imaging allows for exploration and manipulation of data.

\end{abstract}

\keywords{exoplanets, high contrast imaging, integral field spectroscopy, polarimetry, data processing}

\section{INTRODUCTION}
\label{sec:intro}  
This is the first paper in a coordinated series on the calibration and processing of observations from the Gemini Planet Imager (GPI). Ten papers in this series are presented at this meeting. The algorithm performance and calibrations described herein are based on results achieved on sky during GPI's first light run in November 2013 and three subsequent commissioning runs from  December 2013 to May 2014, as well as laboratory testing during GPI's development in California and Canada, and in the instrument lab at Gemini South. These papers seek to describe our current understanding of how best to transform raw data from GPI into quantitative knowledge of nearby planetary systems and other astrophysical observables, based on our experiences working with the instrument for its first six months on sky.  These methods will undoubtedly continue to evolve as GPI operations further mature over the years to come.

GPI is a new facility-class instrument at Gemini Observatory dedicated to directly imaging and spectroscopically characterizing extrasolar planets \cite{Macintosh2008,Macintosh2014}. It combines a very high order adaptive optics (AO) system\cite{Poyneer2008}, a diffraction-suppressing coronagraph\cite{Soummer2011}, and an integral field spectrograph with low spectral resolution but high spatial resolution. GPI was developed by a consortium of institutions spread across North America. Its integral field spectrograph\cite{Chilcote2012SPIE,Larkin2014SPIE}  (IFS) and data pipeline\cite{Maire2010SPIE,Maire2012SPIE} were designed and implemented in a partnership between groups at UCLA and Universit\'e de Montr\'eal; the data pipeline and calibration effort has since expanded to involve collaborators across all GPI partner institutions and is currently coordinated from STScI. Integration and testing of GPI as a whole and its associated software took place at the University of California, Santa Cruz during 2010-2013, leading up to GPI's shipment to Gemini South in August 2013. 

First light of GPI occured on 2013 Nov 12 UT. The instrument performed well from its debut, achieving sufficient contrast during the first light run for imaging detection of $\beta$~Pictoris~b in single 60 s exposures without any point spread function (PSF) subtraction required. See Macintosh et al. 2014 (\refnum{Macintosh2014}) for more details on instrument design, first light performance, and initial science results. As of this writing, five roughly week-long observing runs have been completed including four first light and commissioning runs and one shared risk ``early science'' period. We have acquired data covering a wide range of astronomical science targets and calibration standards in all of GPI's observational modes.  Portions of this data have already been released publicly, and all GPI data will eventually be made public through the Gemini Science Archive.  

The Gemini Planet Imager Data Reduction Pipeline (hereafter GPI DRP or ``the pipeline'') software subsystem allows transformation of raw GPI data into calibrated spectral and polarimetric data cubes. Integral field spectrographs are scientificially powerful but require sophisticated data reduction systems.\cite{2006SPIE.6269E..42L,2006NewAR..50..398A,2010SPIE.7737E..42M,2011PASP..123..746Z,2012AN....333..101B}. 
While $\beta$~Pictoris~b is sufficiently bright to be easily seen in completely raw GPI IFS data (indeed, even without transforming 2D raw images into spectral datacubes), this will not be the case for the vast majority of planets to be seen by GPI.  Before fainter planets can be seen, substantial processing is needed for assembling tens of thousands of microspectra into calibrated spectral datacubes, removing instrumental artifacts, and performing PSF subtractions to further enhance contrast---or equally for GPI's polarimetric mode, reassembling tens of thousands of pairs of dispersed linearly polarized spots and deriving Stokes vector datacubes to gain sensitivity to faint circumstellar dust.

This processing is achieved through the GPI DRP developed by our team and made available to the community. This software may be downloaded from the GPI instrument team's web site\footnote{http://planetimager.org/datapipeline}. A comprehensive software manual is also available online at the same site.  
Previous papers described the early development of the GPI data pipeline and data simulator\cite{Maire2010SPIE} and laboratory test results\cite{Maire2012SPIE}.
This current paper provides an overview of the GPI pipeline as of GPI's first light, describing its requirements, software architecture, and the design choices made in developing it; it is not a substitute for the detailed software manual and we advise users of the instrument to consult that manual for detailed guidance in the use of the pipeline. Based on feedback from observers following the GPI ``early science'' shared-risk observing program, the GPI pipeline and its documentation are already sufficiently mature and user-friendly to enable users with no prior GPI observing experience to rapidly achieve science-ready and preliminarily calibrated datacubes. However, some aspects of calibration are subtle, and in our experience the most time-consuming and critical step in working with GPI data is developing a detailed understanding of systematic noise sources and biases.  Advice on best practices for calibrating GPI data is also available in an IFS Data Handbook on the team's web site\footnote{http://docs.planetimager.org/pipeline/ifs/index.html}, as well as this series of papers.

\section{OUTLINE OF THIS SERIES} 

Subsequent papers in this series discuss in more detail specific aspects of processing, calibrating, and analyzing GPI observations. The series contents are as follows:

\begin{itemize}
\setlength{\itemsep}{1pt}
\item \textbf{Paper I:} Overview of the GPI data analysis pipeline (This work)
\item \textbf{Paper II:} Detector Performance and Calibration (Ingraham et al., paper \#9147-286)
\item \textbf{Paper III:} Empirical Measurement and Applications of High-Resolution Microlens PSFs\\ (Ingraham et al., \#9147-282)
\item \textbf{Paper IV:} Wavelength Calibration and Flexure Correction for the IFS (Wolff et al., \#9147-279) 
\item \textbf{Paper V:} Astrometry and Distortion (Konopacky et al., \#9147-306)
\item \textbf{Paper VI:} Photometric and Spectroscopic Calibration for the IFS (Maire et al., \#9147-307)
\item \textbf{Paper VII:} On-sky Polarimetric Performance  (Wiktorowicz, Max et al., \#9147-305)
\item \textbf{Paper VIII:} Characterization and Role of the Satellite Spots (Wang et al., \#9147-195)
\item \textbf{Paper IX:} Least Square Inversion Flux Extraction (Draper et al., \#9147-189) 
\item \textbf{Paper X:} Non-Redundant Masking on GPI   (Greenbaum et al., \#9147-135) 
\end{itemize}

In addition we encourage interested readers to consult the many other papers on GPI from this meeting, in particular those on the instrument overall, the integral field spectrograph, AO performance, and PSF subtraction.

\section{DATA PIPELINE SOFTWARE ARCHITECTURE} 

\subsection{Context and Requirements}

The GPI data pipeline is required to work in at least three different environments: It must be capable of running in an entirely automated fashion at the telescope to produce preliminary ``quick look'' reduced files for telescope operators and observers. It must be capable of running in an interactive fashion on any Gemini user's computers to allow the production of calibrated science-ready datasets, including providing flexible options to allow end users to adapt the pipeline to the needs of their particular science goals. And it must be capable of being integrated into higher-level data systems for large projects, particularly the Gemini Planet Imager Exoplanet Survey (GPIES) campaign.    The same software executes in all three contexts, though the specific steps included in reduction recipes will differ. Given the diversity of the Gemini community, the pipeline is required to be cross-platform compatible across Mac OS, Windows, and Linux, and it must be operable without purchasing any commercial software licenses (although not necessarily modifiable in that case). 

The most critical focus for the pipeline is transforming raw data into data cubes with minimal instrumental and processing artifacts, in both polarization and spectral modes, and for both coronagraphic and non-coronagraphic observations.  PSF subtraction is a secondary focus; we assume many users will wish to use their own PSF subtraction algorithms, although we do provide moderately sophisticated tools for doing so in the pipeline.

\subsection{Major components}

\begin{figure}[t]
\begin{center}
    \includegraphics[width=0.8\textwidth]{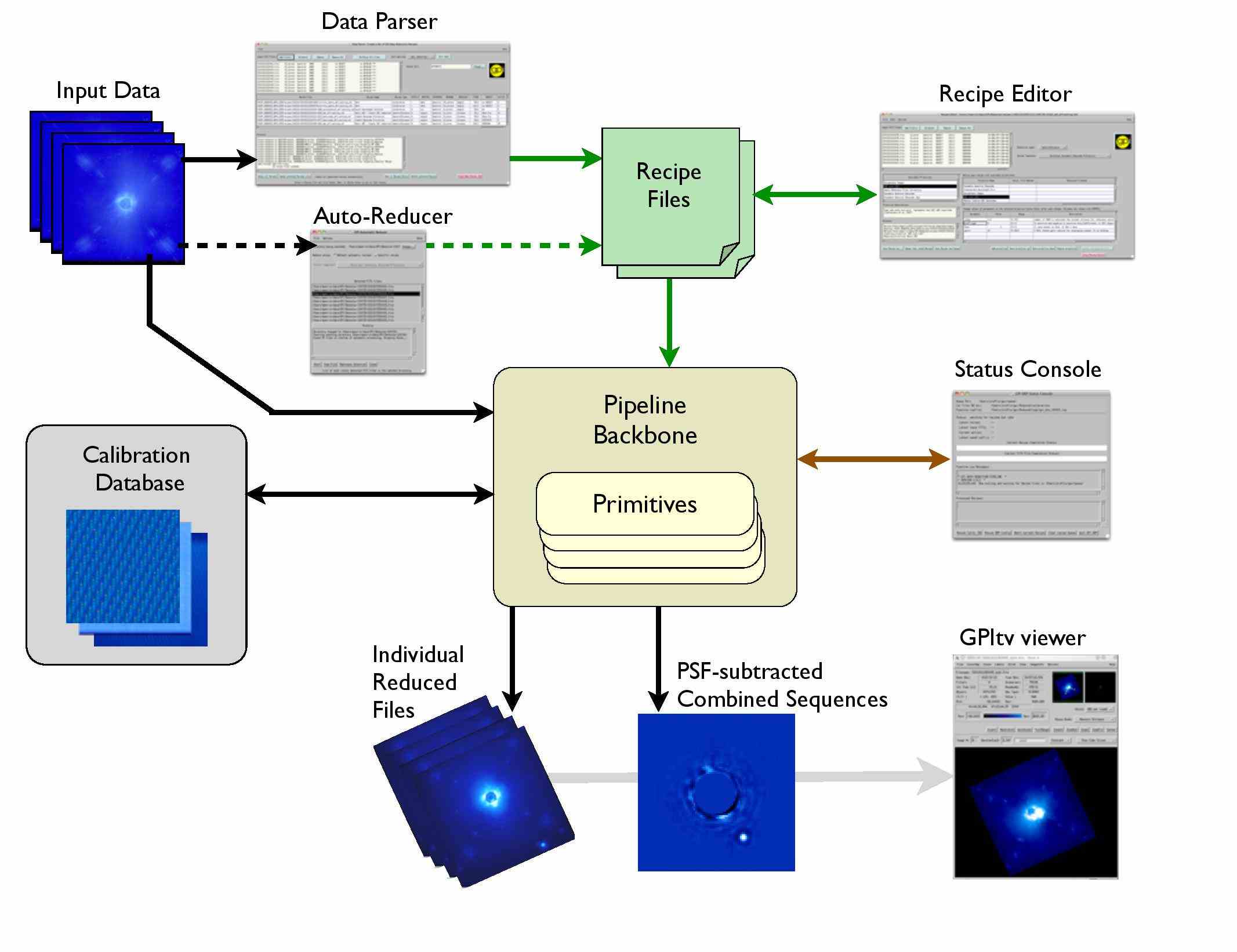}
\end{center}
\caption{Schematic of the components of the GPI data pipeline and data
    flow between them. Reduction recipes are created by the Data
    Parser, or by the Auto-reducer (primarily used at Gemini South
    itself), and recipes may be edited or new recipes created
    interactively using the Recipe Editor. The pipeline backbone
    runs these recipes by executing one or more of a large set of
    reduction primitives. Output files include reduced calibration
    files, individual reduced science files, and reduced combined
    sequences, which can then be examined using the GPItv viewer tool
    or any other FITS datacube compatible viewer. Black and grey connecting lines 
    indicate FITS files, green lines indicate recipe files, and brown indicates 
    status information and some user commands.
}
\label{fig:components}
\end{figure}

The data pipeline system comprises several main components. See Figure \ref{fig:components} for a schematic diagram of the connections between these components.
\begin{itemize}
\item The \textbf{Data Reduction Pipeline core software} (``backbone'') itself, which is responsible for executing the desired reduction sequences. Reduction tasks are defined by \textit{Recipes}, which list both the specific reduction steps (\textit{Primitives}) and the data files to which those steps should be applied. When running, the DRP constantly monitors a queue for new recipes to execute. 
\item \textbf{Graphical user interfaces} for controlling pipeline operation. During observations, an \textit{Autoreducer} generates quicklook reduction recipes on the fly as data is taken. After observations, the \textit{Data Parser} and \textit{Recipe Editor} tools can be used to create and adjust additional recipes. A \textit{Status Console} provides a status display of the data reduction process, keeps a log of all messages generated during reduction, and provides a top-level user console for tasks such as reloading pipeline configuration or cancelling an executing reduction.
\item A \textbf{Calibration Database} that stores reduced calibration files of various types, and automatically provides appropriate files as needed during each step of data processing. Newly created calibration files are stored in the Calibrations Database directory and automatically indexed by the pipeline.
\item And a customized \textbf{data viewer \& analysis tool} called \textit{GPItv}, which allows interactive exploration of images and spectra optimized for GPI data cubes. The pipeline can also produce a variety of other graphical ouputs as plots on screen or files saved to disk. 
\end{itemize}

A reduction recipe consists of a series of some number of data processing primitive steps, a list of one or more input files to operate on, and some ancillary information such as what directory the output files should be written to and a descriptive name for the recipe. GPI recipes are saved as XML files, and while they may be edited by hand, they are more easily created through the use of the Recipe Editor and Data Parser tools. Predefined \textit{Recipe Templates} exist for standard GPI reduction tasks, such as creating wavecals or reducing a series of coronagraphic observations. These templates can be selected and applied to data using the GUI tools. The Data Parser works by examining all files in a given directory and identifying sequences of related files based on FITS header information including instrument configuration, target name, Gemini program ID and datalabel, etc. It has predefined rules for which recipe template is most appropriate for each kind of observing sequence, and the overall order in which recipes should be executed (calibration reductions before science reductions, and so on). Users may view and edit these recipes before running them.

The most complex step in the data reduction is the conversion from 2D raw IFS frames to 3D datacubes. Because the light from each lenslet in the IFS is dispersed across many detector pixels, the process of uniquely assigning flux from detector pixels back to GPI's field of view is non-trivial. This is called ``Datacube Assembly''. The overall process is analogous for both spectral and polarimetry modes, though algorithmic details differ.

The DRP includes primitives for PSF subtraction using either Spectral or Angular Differential Imaging, implemented using both least-squares (Locally Optimized Combination of Images; LOCI\cite{Lafreniere2007}) and principal component analysis (Karhunen-Lo{\`e}ve Image Projection; KLIP\cite{Soummer2012ApJ}) algorithms. But it is expected that many users will wish to use PSF subtraction methods of their own devising on the output data cubes from the pipeline. Likewise the GPI DRP includes primitives for polarimetric differential imaging\cite{Perrin2010SPIE}.

\subsection{Software Infrastructure and Implementation}

The GPI pipeline is implemented in the IDL computer programming language (Exelis Visual Information Solutions, Boulder, Colorado). This was chosen circa 2005 early in the development of GPI, as a more modern development environment than IRAF\footnote{Were we to make this decision today, the pipeline would likely be written in Python, but the Python astronomy ecosystem was much less mature when GPI pipeline development began. Indeed, early decisions about GPI pipeline development were made well before Python's \texttt{numpy} package itself was first released. Cautionary tale: the development timescale for major instrumentation projects is now sufficiently long that the software landscape at a project's completion will likely be radically different than at its inception.}. IDL is a commercial product from Exelis Visual Information Solutions that requires license fees for development; however it is possible to compile IDL code to run in a free-to-distribute IDL Virtual Machine. The GPI pipeline is designed for compatibility with the IDL Virtual Machine, and compiled executables are available for Mac OS, Linux, and Windows platforms from the GPI team website along with the source code.  The pipeline source is compatible with IDL versions $\geq 7$. 

Given the complexity of GPI data reduction tasks, graphical user
tools are essential to aid the user in processing data efficiently, iteratively adjusting recipes if needed, viewing results, and so on. 
We developed a multiprocess archicture using two IDL sessions where
one process runs the GUIs and the other runs the calculations, so that
the GUIs remain responsive even during long calculations. The two IDL
sessions communicate via shared memory buffers, allowing commands and
data to be passed seamlessly between the two.

The pipeline is designed to be relatively straightforward to install and configure, using standard conventions based on environment variables and an optional text configuration file in each user's home directory. Rather than using a separate relational database (e.g., SQLite or equivalent) the calibration database is implemented entirely in IDL as a simple lookup table and associated functions for updates and queries.  The queue directory mechanism is implemented simply using a directory on disk with recipe processing status flagged in filenames, rather than using any more complicated construct such as a FIFO buffer (named pipe) or other message queue. This is intentional simplicity to enable cross-platform deployment and installation on any astronomer's computer with a minimum amount of setup work.

Output images and datacubes are saved in Gemini-standard multi extension FITS format; some derived products such as contrast measurements or extracted point source
spectra may be saved as ASCII tables as well as FITS tables. Datacubes
are written with World Coordinate System (WCS) standard\cite{Greisen2002} compatible FITS headers in both 
spectral and polarimetric modes. 

In the Gemini South control room, the pipeline runs in automated mode to produce ``quick look'' quality preliminary datacubes. These are displayed on screen for inspection by observers in near real time, typically within 10 s of each new raw file becoming available on disk. 
The main purpose of the quick look processing is to allow the observer to monitor instrument performance and make sure that the desired raw contrast  (i.e. before PSF subtraction) is reached.
The quicklook reduction recipes are intentionally simple and do not include all the steps needed for fully phometrically and astrometrically calibrated datacubes, and they do not combine multiple exposures in a sequence.
It is planned that these quicklook files will eventually be incorporated into
the Gemini Science Archive and made available to queue observers but this is
not yet activated at the time of this writing.  

Observers may then use their
own copy of the pipeline to produce more comprehensively calibrated
science-ready products. Feedback from users during and following the GPI Early Science shared-risk observing run 
has generally been positive: The GPI pipeline software and its associated documentation allow observers with zero prior experience with GPI to 
successfully reduce GPI observations obtained in queue mode. However, developing a comprehensive understanding of systematics and performance limitations
is an ongoing task even for the instrument team, and will likely remain so for some time to come; we encourage users to exercise appropriate caution and skepticism when dealing with the data from this new and complex instrument.

\subsection{Heritage from Other Projects}

The GPI pipeline software architecture was in part inspired by, and portions of the backbone code re-used from, the Keck OSIRIS pipeline\cite{2004SPIE.5492.1403K,2004SPIE.5496..426W}. The Recipe Editor was inspired by the OSIRIS \texttt{ODRFGUI}.  GPItv is based originally on \texttt{atv} by Aaron Barth et al.\cite{2001ASPC..238..385B}, with additional portions of code reused from the OSIRIS Quicklook tool \texttt{ql2} (itself derived from the NIRC2/NIRspec quicklook) and from David Fanning's tool \texttt{selectimage.pro}.

The GPI pipeline is built upon an extensive foundation of other open source software from the astronomical community, including but not limited to the Goddard IDL Astronomy Library, the JHUAPL library, Craig Markwardt's MPFIT library, David Fanning's Coyote Library, and other individual tools and routines. See the GPI pipeline software manual for a detailed list of sources.  To simplify the ease of installation of the GPI pipeline and to ensure version compatibility, copies of all of these dependencies are included directly with the pipeline distribution files.

\begin{figure}
\begin{center}
\includegraphics[width=4.5in]{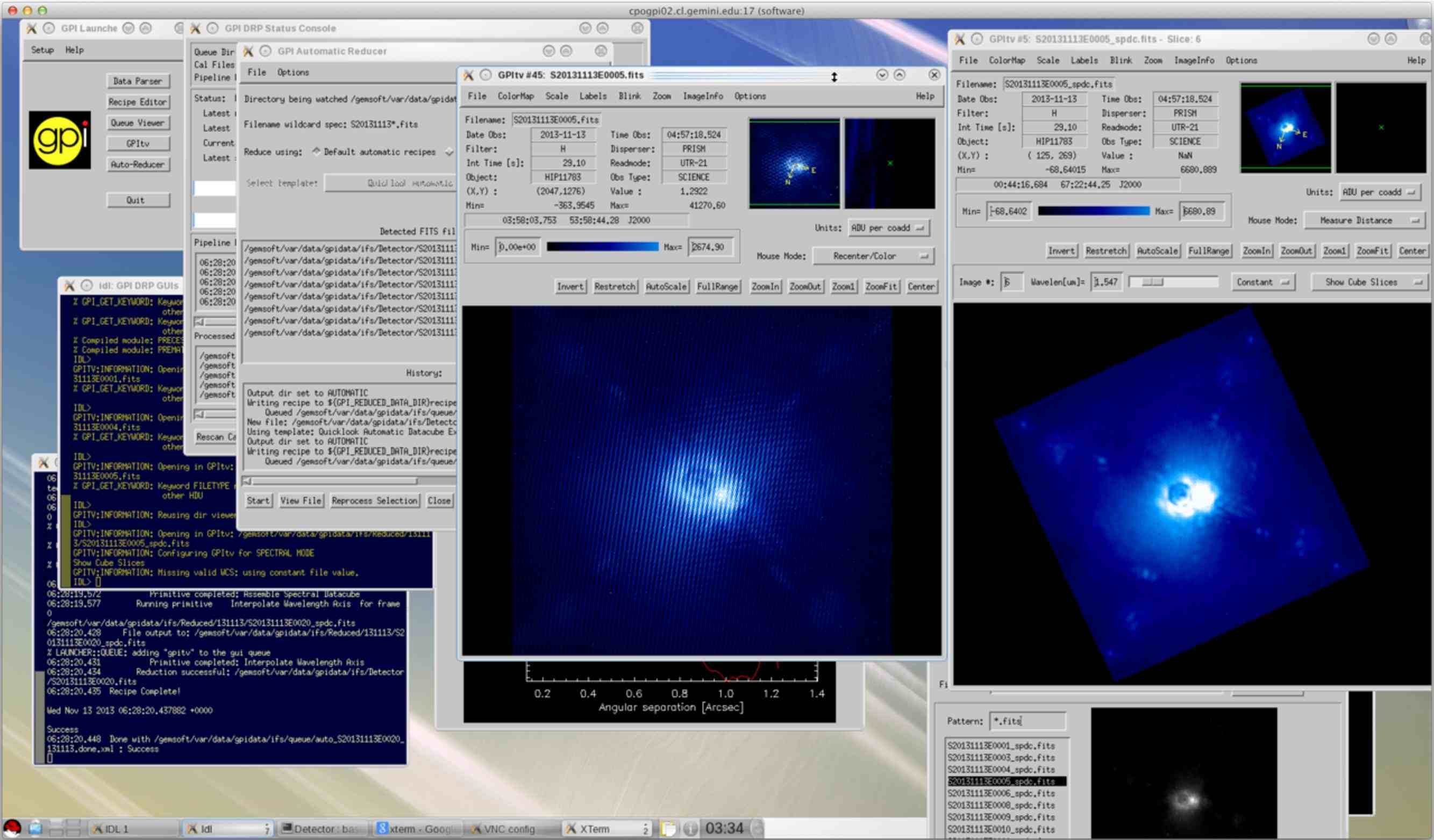}
\includegraphics[width=2in]{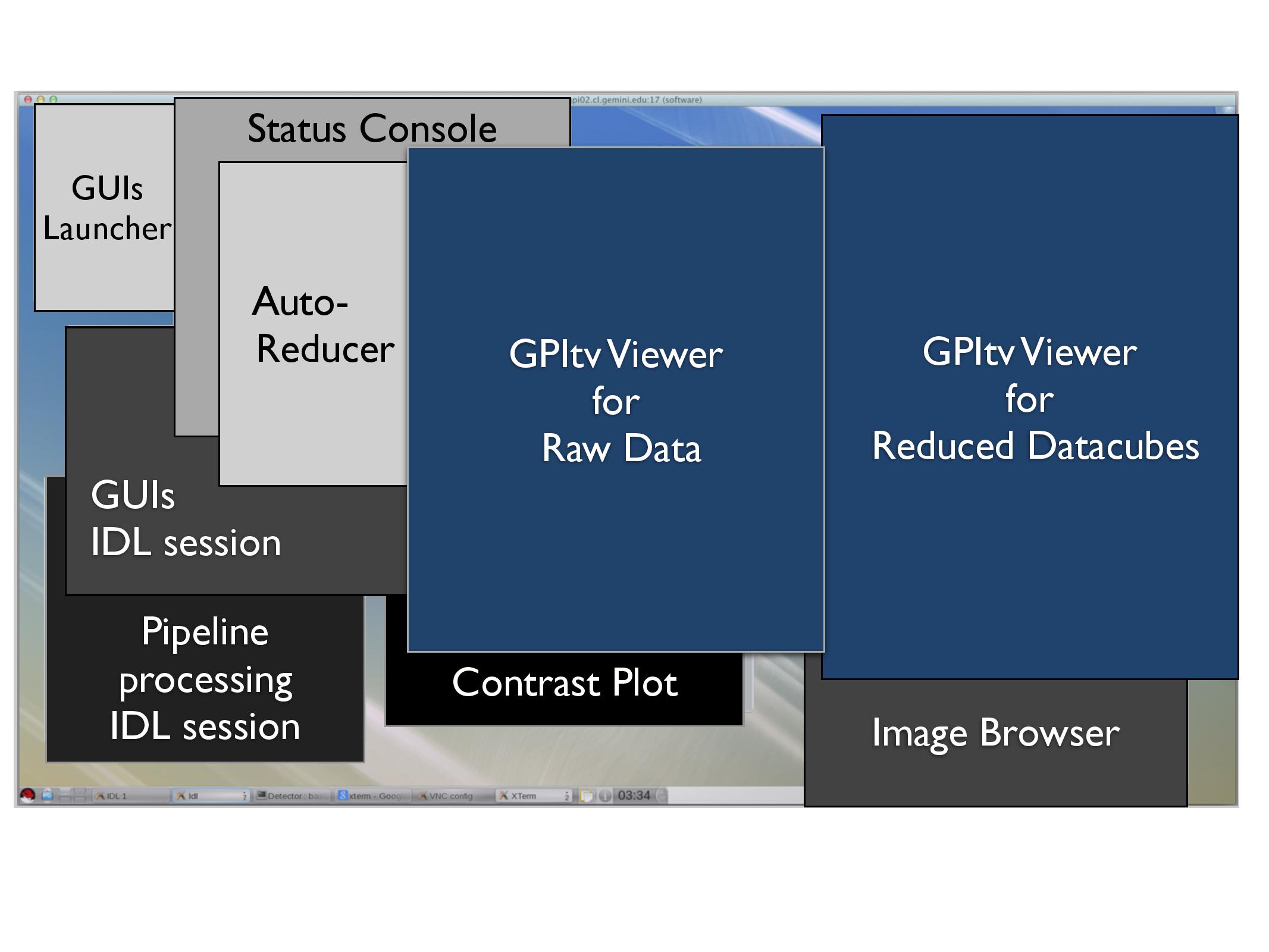}
\end{center}
\caption{Screen shot of the GPI data pipeline running at Gemini South, immediately after first light of the integral field spectrograph and pipeline on 2013 Nov 13 UT. The small panel at right gives an annotated key identifying each of the windows on screen. Two GPItv viewer windows are visible, showing the raw 2D data read from the detector at center and the reconstructed spectral datacube at right. Other windows are the GUI Launcher, from which other tools can be started, the  Status Console window giving overall status and control of the pipeline, and the Auto-reducer which monitors for new files and creates recipes to reduce them. The two terminal windows show the IDL session command lines where diagnostic and log messages are displayed.}
\label{fig:screenshot_firstlight}
\end{figure}

\subsection{Project Management}

The pipeline has passed formal verification tests both as an individual subsystem\cite{Maire2012SPIE} and after integration with the rest of the instrument as part of the GPI pre-ship review process.  The GPI pipeline development team uses a Subversion\footnote{http://subversion.apache.org} repository for source version control, and maintains an issue tracker database using Redmine\footnote{http://www.redmine.org} to
manage bugs and new feature requests.  Documentation is built with Sphinx\footnote{http://www.sphinx-doc.org} and made available in HTML format. Dropbox\footnote{http://www.dropbox.com} is used to share large observational datasets among team members, and we make extensive use of Google Drive\footnote{http://drive.google.com} shared documents.  A weekly GPI data team telecon serves as a venue to coordinate
activities both for pipeline development and for overall instrument
calibration. Support for other observers using GPI may be provided in part by Gemini Observatory and 
in part by some members of the GPI instrument team on a volunteer basis.

The GPI Data Reduction Pipeline is released under a Berkeley Standard Distribution (BSD) 3-clause open source license\footnote{http://docs.planetimager.org/pipeline/license.html}; the included library dependencies are distributed each under its own particular license. 

Not counting those dependencies, the GPI data pipeline is $\sim$80,000 lines of IDL code. It is difficult to estimate the total effort that went into its development, split across many team members working collaboratively, but estimates of 10-20 person-years seem plausible for pipeline software development alone, plus probably several times that much devoted to closely related instrument calibration and data analysis tasks, plus the development of the data simulators for the IFS and AO system used for pipeline development prior to the availability of the real instrument.

\section{SUMMARY OF STEPS IN PROCESSING GPI DATA}

We briefly summarize the major steps in processing GPI data. Further details may be found in the subsequent papers in this series. This is not a comprehensive summary of all available pipeline primitives and their functionality; see the pipeline documentation online for additional information and updates.

\begin{figure}
\begin{center}
\includegraphics[width=6in]{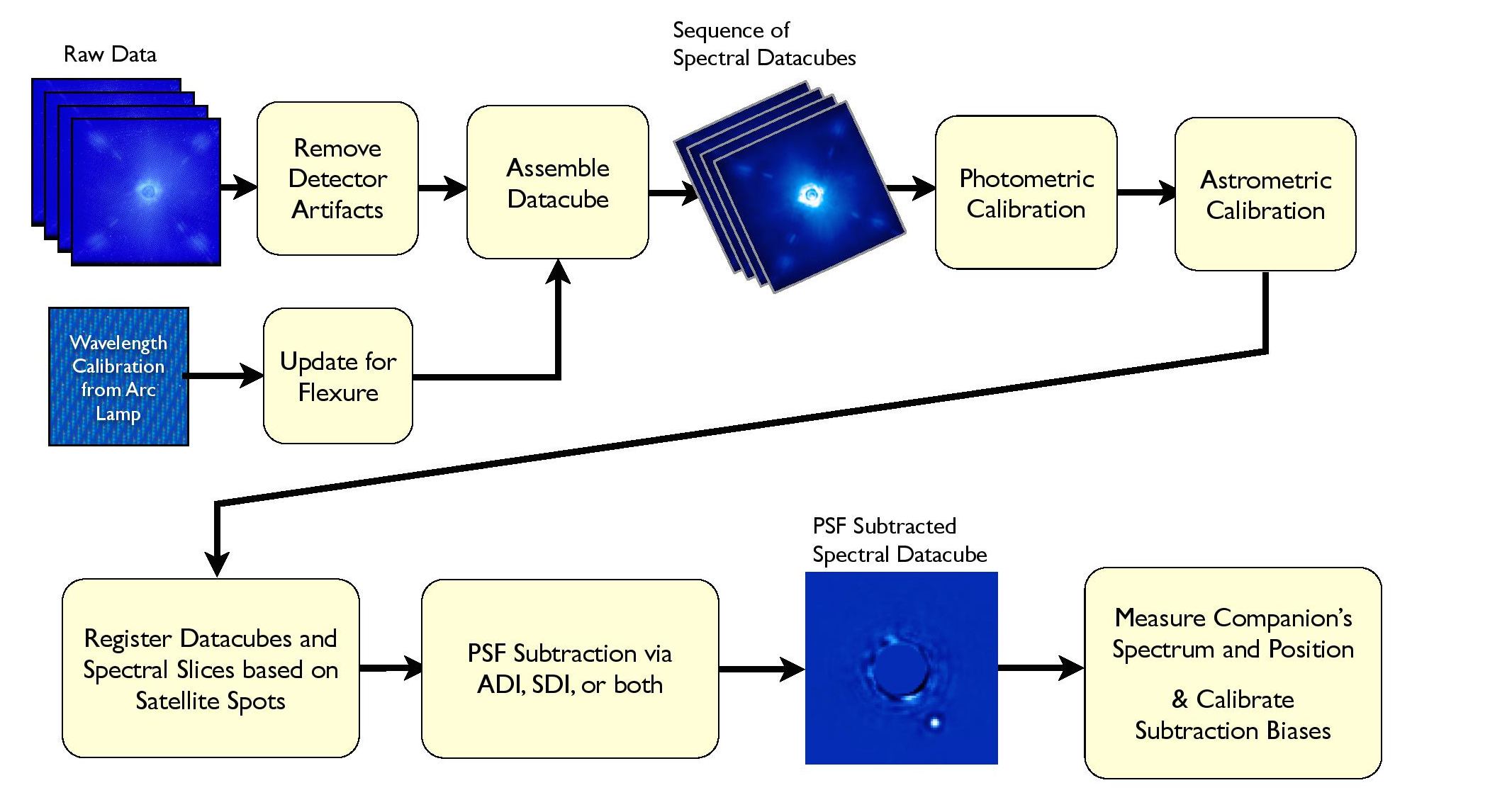}
\end{center}
\caption{Highly simplified diagram of the major steps in processing GPI data for spectral mode. See the text and the following papers in this series for more details on the individual processing steps. }
\label{steps}
\end{figure}

\subsection{Detector calibrations}
Starting from raw files written by the instrument, the process begins with removal of detector systematics (Paper~II). Dark frames are subtracted to remove the background.
This is followed by a ``destriping'' step to remove correlated noise due to the readout electronics and microphonics induced in the H2RG detector by mechanical vibration from the IFS cryocoolers\cite{Chilcote2012SPIE}. Using maps of bad pixels previously generated from calibration observations these pixels are identified and replaced with interpolated values. The Data Quality extension is updated accordingly to track these pixels.
Persistence from prior exposures may be modeled and subtracted if needed. 

For $K$-band observations only, it is beneficial to subtract an observed background frame to remove thermal emission from the instrument, telescope, and sky. Sky backgrounds can be subtracted for all filters if needed, although for coronagraphic observations this is generally unnecessary as the sky background will be automatically removed as part of PSF subtraction. 

\subsection{Spectral mode}

To assemble a datacube, we must know the
locations of the tens of thousands of dispersed lenslet images. Their
pattern is close to but not perfectly uniform due to chromatic distortions in the
spectrograph optics, nonuniformities in the lenslet array itself, etc. Similarly, the individual PSFs of the lenslets vary
smoothly across the field of view, due to the intrinsic and as-built optical aberrations in the refractive spectrograph collimator and camera. 
The properties of these lenslet microPSFs have been empirically calibrated at subpixel resolution (Paper III) using methods adapted from 
high precision point spread function modeling for precision astrometry with the {\it Hubble Space Telescope}. 

In spectral mode the process of mapping the locations of the lenslets occurs as part of wavelength calibration (Paper IV).
We make use of the Gemini GCAL facility calibration system's Xe and Ar emission lamps to fit models of detector pixel position versus wavelength for all lenslets. 
The resulting ``wavecal'' file gives a map of the transformation between datacube coordinates ($X_L$, $Y_L$, $\lambda$) and detector coordinates ($x_d$, $y_d$).  
As GPI points across the sky and the spectrograph changes in elevation, 
internal flexure in the spectrograph optics  lead to small shifts of the spectral positions, with typical magnitude of 0.1-1 pixel and occasionally more. 
Hysteresis and nonrepeatability mean these shifts are only approximately predictable, and must be either measured on an arc lamp image taken at similar telescope attitude to the science data (Paper IV) or derived via forward modeling of the overall lenslet pattern in the science data itself (Paper IX). 

Using a wavecal updated for flexure, the spectral data cube may then be assembled. We currently use a relatively simple but robust and computationally fast method, based on a
 moving box extraction aperture 3$\times$1 pixel in size.  More sophisticated methods are also under development (Paper IX). 
A flat field correction for variations in lenslet throughput can then
be applied. Residual bad pixels may be   detected as local statistical
outliers, and their values replaced through interpolation to yield cosmetically
clean images.

The astrometric properties of GPI have been calibrated based on both lab and on-sky tests, to enable accurate measurement of the positions of astronomical sources (Paper V).
Photometric and spectroscopic calibrations based on standard stars enable reconstruction of source spectra (Paper VI). A pattern of ``satellite spots'', intentional ghost images of the occulted star created by a pupil plane square diffractive grid on the coronagraphic apodizer, allows astrometric and photometric measurements relative to the central star even once it has been blocked from direct view (Paper VIII). 

\subsection{Spectral mode PSF subtractions}

The pipeline provides primitives for PSF subtraction using either spectral differential imaging (SDI) or angular differential imaging (ADI), and these primitives may be chained together to perform both ADI and SDI on the same dataset (though as two steps rather than a unified process). Specifically the pipeline includes impementations of ADI using the original ``classical ADI'' algorithm\cite{Marois2006ApJ}, the LOCI algorithm\cite{Lafreniere2007}, and the KLIP algorithm\cite{Soummer2012ApJ}. The satellite spots are used to locate the star center and register the images together prior to any of these subtraction methods.   That said, as noted above we expect many users will wish to perform their own PSF subtractions using alternate algorithms, so these offerings are not intended to be comprehensive. Similarly, users are responsible for calibrating and understanding the algorithm throughput (i.e. self subtraction bias) of their chosen PSF subtraction method, though we do provide a primitive for injecting mock companions into datacubes for testing.

\subsection{Polarimetric mode}

In polarization mode, each lenslet is dispersed to two orthogonally linearly polarized broadband spots. 
A map of all spot locations is made by fitting model lenslet PSFs to a high S/N polarimetry mode observation of the GCAL quartz halogen flat
lamp. The  resulting spot position calibration file is called a ``polcal'' file.
To assemble polarization pair datacubes, the flux in each pol spot is summed over a small aperture. Currently this
uses a fixed $5\times5$ pixel box, but future versions of the pipeline will
perform optimal extraction using polarized versions of the lenslet PSF empirical models.

A unique aspect of GPI polarimetry data is that, unlike prior instruments, polarimetry observations are taken in ADI mode which complicates the derivation of Stokes parameters because the sky projected polarization angles are constantly changing in the instrument frame. Polarimetric sequence are therefore combined using a forward modeling approach that takes
into account the sky rotation as well as the half wave plate modulation. Differential polarimetry yields improved contrast for polarized scattered light around stars, while the same data can be processed using ADI recipes for PSF subtraction in total intensity.  Further details on reduction algorithms for GPI polarimetry are available in \refnum{Perrin2014PolHR4796A}.

Polarimetric performance on sky, including both speckle suppression and polarimetric accuracy, has been evaluated based on observations of unpolarized and polarized standard stars (Paper VII). 
These data also yield measurements of GPI's instrumental polarization for use in calibration of science data.

\subsection{Non-redundant aperture masking}

GPI also includes a non-redundant mask (NRM) mode\cite{2010SPIE.7735E.266S} for work separations as small as $\lambda/2D$. The performance of GPI NRM data has been evaluated on sky by observations of point source calibrators and resolved targets (Paper X). NRM observations are possible in both spectral and polarmetric modes. The GPI pipeline does not include any special facilities for analysis of NRM data, but is used to produce calibrated datacubes (similarly as for any other observation) for subsequent analysis using other NRM-specific tools.

\subsection{Spatially resolved Solar System targets}
\begin{figure}
\begin{center}
\includegraphics[width=6in]{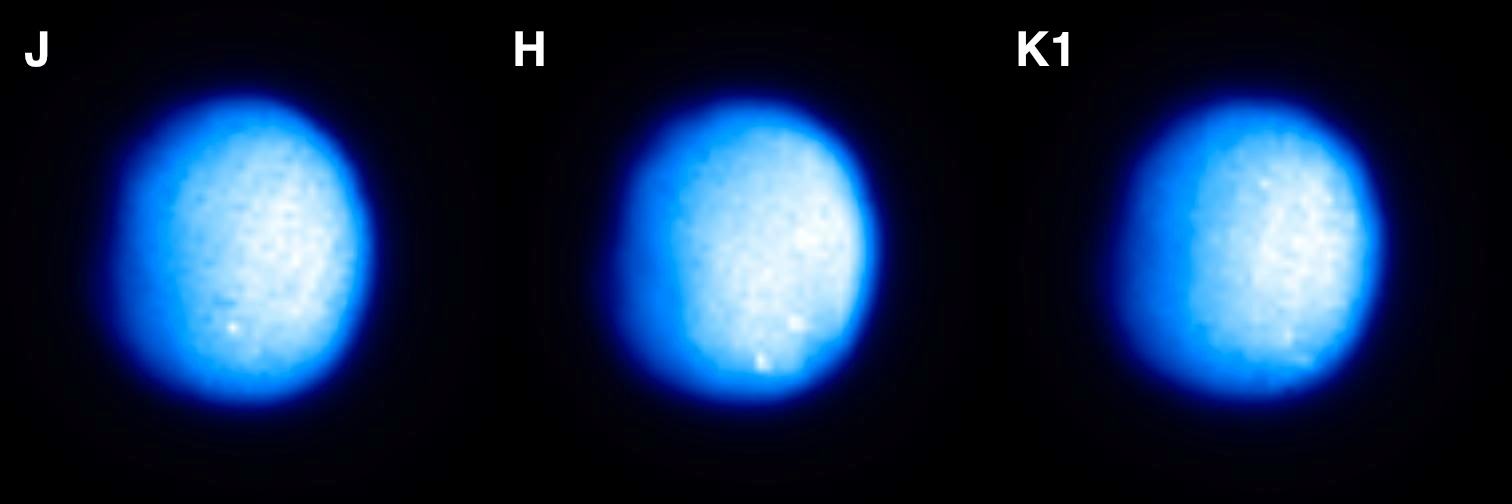}
\end{center}
\caption{Asteroid (2) Pallas was observed with GPI on March 22 2014 in the $J$, $H$ and $K1$ filters (4 min.\ total exposure per filter). These images are the broadband combination of the individual spectral datacubes, after shifting to compensate for atmospheric differential refraction.  North is up, East is left. The 540-km asteroid is well-resolved and irregular. Its silhouette is fitted by an ellipse of 540$\pm$9 mas and 470$\pm$9 mas.
The surface of the asteroid is mostly featureless, however some vertical striping or moir\'e is visible particularly in $J$ band, and there are a handful of residual bright spots from hot pixels. Our experience thus far is that solar system observations offer
a particularly nice test case for validating data pipeline performance and measuring residual systematics, in addition to their own intrinsic scientific value.  \label{fig:pallas} }
\end{figure}

Despite the need for a bright reference source ($I<$10.5), GPI is also capable of observing several solar system targets including Neptune, the Galilean satellites, Titan, and $\sim$50 main-belt asteroids. For instance, the large asteroid Pallas was observed during the March 2014 GPI commissioning run. See Fig.~\ref{fig:pallas} for pipeline processed images. In some ways solar system observations are particularly stringent tests of data pipeline performance, being extended objects with spatially varying surface brightnesses. Imperfections in data cube extraction are readily seen here, especially since these data are not processed via the differential PSF subtraction techniques that naturally subtract out many types of residual artifacts from high contrast data.   
Solar system observations with GPI will be useful for a variety of studies, for instance measurements of the shape of those bodies, mapping surface compositions, monitoring the variability of Titan's atmosphere or volcanoes on Io, and searching for small companion of asteroids\cite{2005Natur.436..822M,2006JGRE..111.7S05D,2013Icar..226.1045H
}.

\section{FUTURE PLANS}
Six months after first light, and six years after software development started in earnest, the GPI data reduction pipeline is working well, enabling a wide range of science observations with GPI by both the instrument team and community observers. Version 1.0 was released in February 2014 to accompany the GPI first public data release, and Version 1.1 in April to support the GPI early science run. The GPI instrument team plans to continue pipeline development for our own use as we carry out the GPI Exoplanet Survey over the next several years.
Particularly high priorities are improving the datacube assembly process making use of the newly available microlens PSF calibrations, and developing better understanding of systematic biases in extracted companion photometry and astrometry. After completion of the Gemini Planet Imager Exoplanet Survey in $\sim$ 2017, we will release a comprehensive archive of calibrated, fully reduced high level science products as part of GPI's legacy to the community.

We will periodically release updated revisions of the pipeline, perhaps once per Gemini observing semester. We encourage collaboration from the community in improving it and welcome code submissions, new pipeline primitives, etc. The GPI data pipeline is open source (BSD license) and is available for adaptation for other projects. For instance it is being used as the starting point for a data pipeline for the optical integral field spectrograph PISCES\cite{2014AAS...22334720M}, a lab prototype being built for the JPL High Contrast Imaging Testbed as a precursor to an IFS for the WFIRST-AFTA coronagraph. In this context, the development, integration, and operation of GPI is an important precursor to that of any dedicated future exoplanet imaging space mission. Just as scientific discoveries from GPI will improve our understanding of exoplanets and feed into the science program of future missions, operational experience with GPI will help refine current concepts for implementing high contrast coronagraphic imaging spectroscopy from space.

\vskip 0.3 in

J.M. and M.D.P. initiated the software development of the pipeline starting in 2007, developed the overall architecture and first versions of all pipeline components and many of the primitives. R. Doyon initially led the DRP subsystem, transitioning to M.D.P. starting in 2011. Authors P.I. through M.P.F. wrote portions of the pipeline code, developed algorithms, and conducted observations and analyses. B.M. and J.R.G. provided overall leadership to the GPI project, and J.R.G. developed algorithms for polarimetry reduction. J.E.L. led the development of the IFS, with collaboration from J.C., M.D.P., R. Doyon and other members of the GPI team; see \refnum{Larkin2014SPIE}. S.G. and D.P. kept GPI on track through project management, while K.L. provided Gemini oversight and feedback on the pipeline development. The remaining authors analyzed data,  collaborated in instrument calibrations and scientific analyses, and conducted observations.

\acknowledgments     
The Gemini Observatory is operated by the Association of Universities for Research in Astronomy, Inc., under a cooperative agreement with the NSF on behalf of the Gemini partnership: the National Science Foundation (United States), the National Research Council (Canada), CONICYT (Chile), the Australian Research Council (Australia), Minist\'erio da Ci\'encia, Tecnologia e Inova\c{c}\=ao (Brazil), and Ministerio de Ciencia, Tecnolog\'ia e Innovaci\'on Productiva (Argentina).
We acknowledge financial support of the Gemini Observatory, the NSF Center for Adaptive Optics at UC Santa Cruz (NSF AST-9876783), the NSF (AST-0909188; AST-1211562), NASA Origins (NNX11AD21G; NNX10AH31G), the University of California Office of the President (LFRP-118057), the STScI Director's Discretionary Research Fund (DDRF-D0101.90164), and the Dunlap Institute, University of Toronto. 
M.D.P. was partially supported by a NSF Astronomy and Astrophysics Postdoctoral Fellowship (AST-0702933). J.M. is a Dunlap Fellow at the Dunlap Institute for Astronomy \& Astrophysics, University of Toronto. The Dunlap Institute is funded through an endowment established by the David Dunlap family and the University of Toronto.
Portions of this work were performed under the auspices of the U.S. Department of Energy by Lawrence Livermore National Laboratory under Contract DE-AC52-07NA27344. Other portions were performed under contract with the California Institute of Technology/Jet Propulsion Laboratory funded by NASA through the Sagan Fellowship Program executed by the
NASA Exoplanet Science Institute.  S.G.W and A.Z.G. are supported by the NSF Graduate Research Fellowship Program.
The GPI team makes use of Dropbox for sharing large datasets among team members and thanks Dropbox for sponsoring a team account.
We are indebted to the international team of engineers and scientists who worked to make GPI a reality, and to the broader community of astronomers and software developers who create the open source tools and libraries upon which we have built.

\bibliographystyle{spiebib}   \bibliography{./references}   
\end{document}